\documentclass[aps,prl,twocolumn,showpacs]{revtex4}
 \usepackage{graphics}
 \usepackage{amsmath,amssymb}
 \usepackage{epsfig}
 \usepackage[english]{babel}

\begin{document}

\title{Proposal for a Nuclear Gamma-Ray Laser of Optical Range.}

\author{E.~V.~Tkalya}

\email{tkalya@srd.sinp.msu.ru}

\affiliation{Institute of Nuclear Physics, Moscow State University, Moscow, Russia}

\date{\today}

\begin{abstract}
A possibility of amplification of the 7.6 eV $\gamma$-radiation by the stimulated $\gamma$-emission of the ensemble of the $^{229m}$Th isomeric nuclei in a host dielectric crystal is proved theoretically. This amplification is a result of: 1) the excitation of a large number of the $^{229m}$Th isomers by laser radiation; 2) the creation of the inverse population of nuclear levels in a cooled sample owing to the interaction of thorium nuclei with the crystal electric field or with an external magnetic field; 3) the emissions/absorption of the optical photons by thorium nuclei in the crystal without recoil; 4) the nuclear spin relaxation through the conduction electrons of the metallic covering.
\end{abstract}


\maketitle

Nuclear $\gamma$-ray laser is a known challenge of the modern physics. To succeed in the field, one has to overcome at least two basic problems, namely: to accumulate a sufficient amount of isomeric nuclei in a sample, and to narrow down the emission $\gamma$-ray line to its natural radiative width.  These and other problems, as well as early unsuccessful attempts of their solution are discussed in detail in reviews \cite{Baldwin-97,Rivlin-07} (see also list of reviews in \cite{Baldwin-97}).

Nevertheless, there is a plausible way to overcome all difficulties and to develop a unique $\gamma$-ray laser working on magnetic dipole ($M$1) transition in the optical range between the first excited level ($J_{is}^{\pi}=3/2^+$)  and the ground state ($J_{gr}^{\pi}=5/2^+$) of the $^{229}$Th nucleus.

According to the recent study \cite{Beck-07} the energy, $E_{is}$, of the abnormally low-lying level in $^{229}$Th is 7.6$\pm$0.5 eV. A working compound with an energy gap, $\Delta$, between the top of the valence band and the bottom of the conduction band such that $\Delta>E_{is}$ was found in \cite{Rellergert-10}, where $^{232}$Th nuclei were implanted in LiCaAlF$_6$. There, the Th$^{4+}$ ions substitute for the Ca$^{2+}$ ions and lead to the formation of the Th:LiCaAlF$_6$ alloy with $\Delta\simeq10$ eV.

Thus there is a unique situation for low-energy nuclear spectroscopy in $^{229}$Th:LiCaAlF$_6$. There, a photon is expected to interact directly with the atomic nucleus thereby bypassing the interaction with the electron shells \cite{Tkalya-00-JETPL}, i.e. the internal conversion and electronic bridge \cite{Tkalya-00-PRC,Tkalya-03}. This opens up interesting perspectives and applications: the $\gamma$-ray emission of the excited $^{229}$Th nuclei in the optical range called the ``nuclear light'' \cite{Tkalya-03}; a check of the dependence of the isomeric level half life on the refractive index \cite{Tkalya-00-JETPL,Tkalya-00-PRC}; the nuclear metrological standard of frequency \cite{Tkalya-96} called  the ``nuclear clock'' \cite{Peik-03}; the M\"{o}ssbauer effect in the optical range \cite{Tkalya-96}; the relative effect of the variation of the fine structure constant $e^2$ and the strong interaction parameter $m_q/\Lambda_{QCD}$ \cite{Flambaum-06}; a check of the exponentiality of the decay law \cite{Dykhne-98}. Results obtained in the present work can be used in practice to study the behavior of the nuclear ensemble with the high density of inverse population, the cooperative spontaneous emission \cite{Dicke-54} and related phenomena.

The most important characteristic of the $\gamma$-ray laser is its amplification factor. The amplification coefficient of gamma radiation per unit length, $\chi$, caused by stimulated emission in a medium is given \cite{Baldwin-97,Rivlin-07} by
\begin{equation}
\chi=\frac{\lambda_{is}^2}{2\pi}\frac{\Gamma_{rad}}{\Delta\omega_{tot}}\frac{1}{1+\alpha} \left( n_{is}-\frac{n_{gr}}{g} \right) -\kappa .
\label{eq:AF}
\end{equation}
Here we work in the system of units $\hbar=c=k=1$, $\lambda_{is}=2\pi/E_{is} = 163\pm11$ nm is the wavelength of the isomeric transition, $\Gamma_{rad}$ is the radiative width of the nuclear transition from the isomeric level to the ground state in vacuum, $\Delta\omega_{tot}$ is the total width of the transition line which accounts for all the types of the homogeneous and inhomogeneous broadening, $\alpha$ is the coefficient of the electronic conversion, $n_{gr(is)}$ is the density of nuclei in the ground (isomeric) state, $g=g_{gr}/g_{is}$ where $g_{gr(is)}=2J_{gr(is)}+1$, and $\kappa$ is the linear attenuation coefficient for photons with the wavelength $\lambda_{is}$ in the medium in all processes excluding the resonant transition $gr\rightarrow{}is$ ($\kappa\simeq1$ cm$^{-1}$ \cite{Shiran-04} and it can be reduced). The amplification of the $\gamma$-radiation occurs if $\chi{}>0$.

The radiating width $\Gamma_{rad}$ in Eq.(\ref{eq:AF}) requires the evaluation of the nuclear matrix element of the $M$1 transition $3/2^+(7.6$ eV)$\rightarrow 5/2^+(0.0)$. One can calculate this matrix element using the Alaga rules and experimental data \cite{Bemis-88,Barci-03,Ruchowska-06} for the $M$1 transition $9/2^+(97.13$ keV)$\rightarrow  7/2^+(71.82$ keV) \cite{Dykhne-98_ME}. Recalculation of the experimental data gives respectively the following values for the reduced probability of the required transition in terms of the Weisskopf units   $B_{W.u.}(M1)=0.038$, 0.024, and 0.014. The Coriolis interaction between rotational bands enhances the transition probability by a factor of 1.2-1.3 \cite{Dykhne-98_ME}. So for the preliminary estimation one can use the ``enhanced'' average value $B_{W.u.}(M1)\approx 0.032$. Correspondingly the value of the radiative width is $\Gamma_{rad}\approx3\times10^{-19}$ eV.

The internal conversion of the $3/2^+(7.6$ eV) level is forbidden and $\alpha=0$ in the $^{229}$Th:LiCaAlF$_6$ crystal because $\Delta>E_{is}$. The probability of the electron bridge \cite{Strizhov-91} is negligibly small in the case $\Delta-E_{is}\geq1$ eV \cite{Tkalya-00-JETPL}.

In the $^{229}$Th:LiCaAlF$_6$ crystal the line width $\Delta\omega_{tot}$ of the transition is determined mainly by the interaction of the nuclear magnetic moments \cite{Rellergert-10}. The magnetic moment of the ground state $\mu_{gr}=0.45\mu_N$ \cite{Bemis-88}, where $\mu_N$ is the nuclear magneton. The magnetic moment of the isomeric level $\mu_{is}$ was calculated in \cite{Dykhne-98_ME} on the basis of the available experimental data: $\mu_{is}\simeq-0.08\mu_N$. With these values for $\mu_{gr}$ and $\mu_{is}$ the value for the broadening of spectral line of the isomeric transition can be reduced to 1 kHz \cite{Rellergert-10} and approximated by $\Delta\omega_{tot}\leq7\times10^{-13}$ eV.

Now we consider the two-step creation of the inverse population in the system of the $^{229}$Th nuclei. The first step is obtaining a considerable number of the isomers $^{229m}$Th(3/2$^+$, 7.6 eV) by photoexcitation of the $^{229}$Th nuclei in the transparent target by the vacuum ultraviolet (VUV) laser photons \cite{Tkalya-96,Tkalya-03}. The equations for $n_{is}$ and $n_{gr}$ as functions of irradiation time are given by
\begin{equation}
dn_{is}/dt=\sigma\varphi{}n_{gr}-\Lambda_{is}n_{is}- g\sigma\varphi{}n_{is}=-dn_{gr}/dt.
\label{eq:Eq2}
\end{equation}
Here $\sigma$ is the resonance cross section of nuclear photo excitation by laser radiation with the wavelength $\lambda_L=\lambda_{is}$ and the line width $\Delta\omega_L$: $\sigma=(\lambda_{is}^2/2\pi)\Gamma_{rad}/(g\Delta\omega_L)$,  $\varphi$ is the flux density of laser photons, $\Lambda_{is}$ is the decay constant of the isomeric level. In our case $\Lambda_{is} = \Gamma_{rad} = \ln(2)/T^{is}_{1/2}$, where $T^{is}_{1/2}\approx25$ min is the level half-life (we take 1 for the refractive index (details see in \cite{Tkalya-00-JETPL,Tkalya-00-PRC})). The term $g\sigma\varphi{}n_{is}$ describes the $\gamma$ emission of the isomeric nuclei stimulated by laser photons. Below we will demonstrate that both emission and absorption occur without recoil. Therefore the laser radiation is in resonance with the direct and return nuclear transitions $5/2^+(0.0)\rightleftarrows3/2^+$(7.6 eV).

The solutions of Eq.~(\ref{eq:Eq2}) at $t\rightarrow\infty$ with the initial conditions $n_{gr}(0)=n_0$ and $n_{is}(0)=0$ are $n_{is} = n_0\sigma\varphi / (\Lambda_{is}+(1+g)\sigma\varphi)$,
$n_{gr} = n_0(\Lambda_{is}+\sigma\varphi{}g)/ (\Lambda_{is}+(1+g)\sigma\varphi)$.
We see, that $n_{gr}/n_{is}=\Lambda_{is}/\sigma\varphi+g>1$, because $g>1$. So, it is impossible to obtain $real$ inverse population, i.e. to fulfill the condition $n_{is}-n_{gr}>0$ by this method. The condition $n_{is}-n_{gr}/g>0$ in Eq.~(\ref{eq:AF}) is not fulfilled too.

On the other hand, by laser radiation pumping one can excite a considerable number of nuclei on the isomeric level. As an initial density of the $^{229}$Th nuclei in the Th:LiCaAlF$_6$ crystal we take $n_0=10^{18}$ cm$^{-3}$ \cite{Rellergert-10}, which amounts to $10^{-4}$ of the LiCaAlF$_6$ crystal density. Such a small number of the Th$^{4+}$ ions in the LiCaAlF$_6$ crystal is not expected to change considerably the electronic band structure of the LiCaAlF$_6$ crystal. Let us consider a crystal of $^{229}$Th:LiCaAlF$_6$ which has a cylindrical form satisfying the condition $L\gg{}D$, where $D=0.01$ cm is the specimen diameter and $L$ is its length. (The Fresnel number $F=\pi{}D^2/(4L\lambda_{is})=10$ if $L=0.5$ cm, and $F=1$ if $L=5$ cm.) Then for a VUV laser with the average power $P=30$ mW estimations show that $n_{is}/n_{0}\approx 1/4$ during a period of radiation $\sim{}T^{is}_{1/2}$.

\begin{figure}[]
\epsfig{file=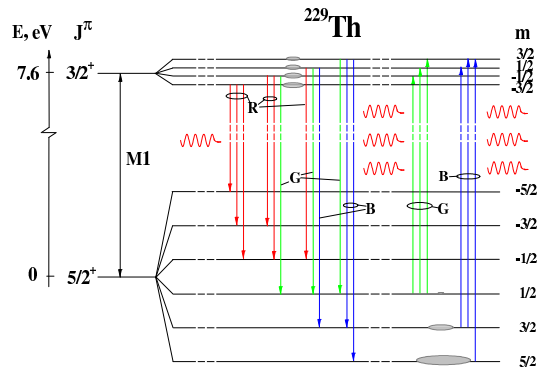,width=7cm}
\caption{The transitions between the Zeeman sublevels of the $^{229}$Th nucleus in magnetic field. The gray ovals are proportional to partial level populations.}
\label{fig:Levels}
\end{figure}

The {\it{effective}} inverse population in the irradiated and cooled sample is facilitated by the magnetic field. The Zeeman splitting of the nuclear levels is shown in Fig.~\ref{fig:Levels}. The spin-lattice relaxation time $T_1$ required to achieve the Boltzmann distribution is unknown in $^{229}$Th:LiCaAlF$_6$ and should be determined experimentally. Let us make tentative assumption that $T_1$ meet the condition $T_1\leq{}T_{1/2}^{is}$. In that case populations of the sublevels are described by  $\exp(-\mu_{gr(is)}\text{H}/T)$. The ratio of the energy differences between the Zeeman sublevels of the ground and isomeric states is approximated by $|\mu_{gr}/\mu_{is}|\approx6$. Therefore the population of the ground state sublevels falls down much faster, than the population of the isomeric sublevels. The magnetic interaction energy $\mu$H is $1.4\times10^{-6}$ eV for the ground state and $2.5\times10^{-7}$ eV for the isomeric state in the magnetic field H = 100 T.
At the temperature $T=0.01$ $^{\circ}$K all sublevels of the isomeric state are populated: $n_{|3/2^+,\text{-}3/2\rangle}$ = $9.5\times10^{16}$ cm$^{-3}$,  $n_{|3/2^+,\text{-}1/2\rangle}$ =$7.1\times10^{16}$ cm$^{-3}$, $n_{|3/2^+, 1/2\rangle}$ = $5.3\times10^{16}$ cm$^{-3}$, and $n_{|3/2^+, 3/2\rangle}$ = $4.0\times10^{16}$ cm$^{-3}$. For the ground state calculation gives
$n_{|5/2^+,5/2\rangle} = 6.0\times10^{17}$ cm$^{-3}$, $n_{|5/2^+,3/2\rangle} = 1.1\times10^{17}$ cm$^{-3}$, and $n_{|5/2^+,1/2\rangle} = 2.1\times10^{16}$ cm$^{-3}$. The population of the other sublevels is negligibly small.

%
%

Thus, the upper Zeeman sublevels of the ground state are unoccupied. Transitions allowed by the selection rules for the $M$1 multipole (they are shown by red (``R'') arrows in Fig.\ref{fig:Levels}) are possible to these states from the sublevels of the isomeric states. Therefore, in the system of nuclear isomers an inverse population is effectively realized. It is worth noting that there is no resonant absorption for the ``red'' photons because their energies are too small to excite the $^{229}$Th nuclei from the populated states $|5/2^+,5/2,3/2,1/2\rangle$.

The relative probability $p$ of all these transitions taking into account the population of the states and the corresponding Clebsch-Gordan coefficients is 0.63. This value is then used for the calculation of the amplification factor, Eq.~\ref{eq:AF}, where we put $n_{gr}=0$ and $pn_{is}$ instead of $n_{is}$. The result is $\chi\simeq3$ cm$^{-1}$.

The transitions shown by three left green (``G'') arrows in Fig. \ref{fig:Levels} also lead to the enhancement of the radiation, but their contribution to $\chi$ is very small ($\simeq1$\%).

The resonant excitation of the isomeric level from the weakly populated state $|5/2^+,1/2\rangle$ is also possible (corresponding transitions are labeled by three right green arrows in Fig. \ref{fig:Levels}). Such an excitation is a result of the M\"{o}ssbauer effect in the optical range. Emission of the $\gamma$-ray photons by the $^{229m}$Th isomers and the resonant absorption of these photons by the $^{229}$Th nuclei in a solid should occur without recoil. Indeed, the energy lost $E_R$ due to the recoil is negligibly small: $E_R=\omega^2/2M\approx 1.4\times10^{-10}$ eV (here $M$ is the $^{229}$Th nucleus mass). The probability of the M\"{o}ssbauer effect at low temperature estimated with the Debye-Waller factor is   $f\approx\exp(-3E_R/2\theta_D)$ \cite{Mossbauer-61}, where $\theta_D$ is the Debye temperature. It is evident that in our case $f=1$ because $E_R/\theta_D\ll 1$. The resonant absorption of the gamma radiation is more pronounced for the remaining transitions shown in Fig. \ref{fig:Levels} by three left and tree right blue (``B'') arrows.

The sign of $\mu_{is}$ (it was predicted theoretically) is not critical for the amplification effect. The positive value of $\mu_{is}$ would simply half the general amplification factor. In such a case the radiation amplification is due to the three transitions to the upper sublevels of the ground state from the upper Zeeman isomeric sublevels.

Let us consider the second scheme of $\gamma$-ray laser which is based on the electric quadrupole splitting. The $^{229}$Th nucleus has appreciable quadrupole moment. The energy of the electric quadrupole interaction $E_{QI}$ is proportional to the product of the nuclear quadrupole moment and the electric field gradient (EFG), $\phi_{zz}$. The ground state spectroscopic quadrupole moment $Q_{gr}$ is 3.15 $e$b \cite{Bemis-88}. If one makes the standard assumption that the intrinsic quadrupole moment $Q_2=8.816$ $e$b \cite{Bemis-88} remains the same for the rotational bands $K=5/2$ and $K=3/2$ \cite{Dykhne-98_ME} then $Q_{is}\approx 1.8$ $e$b. In the $^{229}$Th:LiCaAlF$_6$ crystal the leading contribution to EFG at the Th$^{4+}$ ion site comes from F$^-$ ions, which compensate the extra charge 2$^+$. These ions are located in interstitial sites in the vicinity of Th$^{4+}$ \cite{Amaral-03, Jackson-09}. An estimation gives $\phi_{zz}\simeq -10^{18}$ V cm$^{-2}$ at the Th$^{4+}$ site in $^{229}$Th:LiCaAlF$_6$.

To obtain EFG at the  $^{229}$Th nucleus one has to take into account the effect of antishielding. The Sternheimer antishielding factor, $\gamma_{\infty}$, for Th$^{4+}$ is $\simeq$100--200. This value is extracted from the data approximation \cite{Sen-77} for $\gamma_{\infty}$ in ion series Fr$^{+}$, Ac$^{2+}$, Ra$^{3+}$, which as Th$^{4+}$ have the close electron shell of Rn. Thus, EFG at the $^{229}$Th nucleus  $(1-\gamma_{\infty})\phi_{zz}$ can exceed $-10^{20}$ V cm$^{-2}$. Correspondingly a typical energy of the electric quadrupole interaction $E_{QI}$  can amount to $-10^{-4}$ eV.

It is also worth noting that according to \cite{Rellergert-10}, each Th$^{4+}$ ion occupies in $^{229}$Th:LiCaAlF$_6$ a single lattice site. Therefore, the electric quadrupole interaction results in constant shift of the transition energy, which does not lead to a line broadening.

%
%
\begin{figure}[]
\epsfig{file=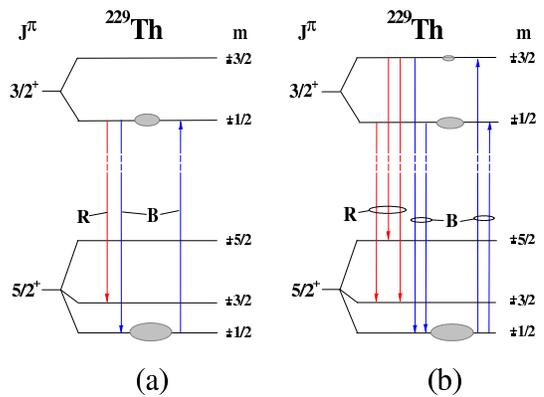,width=7cm}
\caption{The transitions between sublevels of the $^{229}$Th nucleus in the case of quadrupole splitting: (a) the case $T_1\lesssim{}T_{1/2}^{is}$; (b) the case $T_{1/2}^{is}\ll{}T_1$ (see text for details).}
\label{fig:LevelsQ}
\end{figure}

If EFG at the $^{229}$Th nucleus in $^{229}$Th:LiCaAlF$_6$ indeed reaches the abovementioned extreme value, then the nuclear splitting scheme will be as shown in Fig.\ref{fig:LevelsQ}~(a). The sublevels energies are given \cite{Abragam-61} by
$
E_m = eQ_{gr(is)} (1-\gamma_{\infty}) \phi_{zz} (3m^2-J_{gr(is)}(J_{gr(is)}+1)) /
(4J_{gr(is)}(2J_{gr(is)}-1)).
$
At $T=0.01$ $^{\circ}$K it is mainly the sublevels $|5/2^+,\pm1/2\rangle$ ($n=7.4\times10^{17}$ cm$^{-3}$) and $|3/2^+,\pm1/2\rangle$ ($n=2.6\times10^{17}$ cm$^{-3}$) which are populated. The amplifying coefficient for the transition $|3/2^+,\pm1/2\rangle\rightarrow|5/2^+,\pm3/2\rangle$  marked by the red (``R'') arrow in Fig.\ref{fig:LevelsQ}(a) is $\chi\simeq2$ cm$^{-1}$. The blue (``B'') arrows indicate the transitions with the resonant absorption of the gamma radiation. Notice that at large values of EFG the effective inverse population and the amplification condition will hold up to the temperature $T=0.1$ $^{\circ}$K.

Now we consider the third scheme of $\gamma$-ray laser when $T_{1/2}^{is}\ll{}T_1$. In that case the inverse population can be achieved by the following way. At first the crystal is hold at a low temperature for the time $T_1$ required for the Boltzmann distribution population of the sublevels of the ground state of the $^{229}$Th nucleus to occur. Thereafter the laser radiation provides population of the isomeric state. The degree of the crystal heating is governed by the absorption coefficient $\kappa$. Standard purification methods can reduce harmful impurities to part-per-billion levels \cite{Rellergert-10}, which ensures the reduction of $\kappa$ to 0.01 cm$^{-1}$. An estimation shows that an effective laser heat removing even on the scale of few $T_{1/2}^{is}$ will not increase the crystal temperature above $T=0.1$ $^{\circ}$K. When we stop the laser radiation, the cooling down to $T=0.01$ $^{\circ}$K will occur practically instantly. Thus, there will be no redistribution of the nuclear sublevels populations due to laser radiation. The upper sublevels of the $^{229}$Th nuclear ground state will remain barely populated which will lead to final effective inverse population.

The total cross-section for the isomeric excitations by the laser radiation from the $|5/2^+,\pm1/2\rangle$ state is virtually coincides with $\sigma$ in Eq.\ (\ref{eq:Eq2}). Furthermore, 90\% of excited nuclei are in the $|3/2^+,\pm1/2\rangle$ state and 10\% in the $|3/2^+,\pm3/2\rangle$ state. The inverse level population and the amplification will take place at the ``red'' transitions $|3/2^+,\pm1/2\rangle\rightarrow|5/2^+,\pm3/2\rangle$ and $|3/2^+,\pm3/2\rangle\rightarrow|5/2^+,\pm5/2,\pm3/2\rangle$ in Fig.\ref{fig:LevelsQ}(b). The overall amplification factor will be $\chi\simeq3$ cm$^{-1}$.

The duration of the $\gamma$-ray laser emission, $\tau$, can be estimated by the formula $\tau\simeq{} T_{1/2}^{is}(L/D)^2 \exp(-\chi{}L)$ in the case $\tau\ll{}T_{1/2}^{is}$ and $L\lesssim\chi^{-1}\ln(N_{is}/2)$, where $N_{is}=n_{is}(\pi{}D^2/4)L$. For $L=5$ cm $\tau\simeq10^2$ s. The emission will be a sequence of pulses with the repetition frequency $f_{rep}=Q_{is}(D/L)^2\simeq10^{5}$-$10^{4}$ s$^{-1}$, where $Q_{is}=N_{is}\ln(2)/T_{1/2}^{is}$. The mean power of the $\gamma$-ray laser will be $\simeq10^{-7}$ W. For $L\approx1$ cm it is necessary to use surrounding mirrors to ensure sufficient amplification for the light, which makes some passes through the gain medium. In this case $L$ in the formula for $\tau$ should be replaced by a total distance which photons pass inside the sample.

In the following we consider the relaxation time problem in detail. According to theoretical models $T_1$ can range from few minutes up to millions of years in insulators at cryogenic temperatures \cite{Abragam-61}. On the other hand, there exists a process which can substantially reduce $T_1$. The $M1$ transitions between the sublevels can be accelerated of by covering the sample with a non-superconducting metal: Au, Cu, Ag. (Superconductors have the energy gap, which can exceed the level splitting $\Delta{}E$. It forbids the exchange of the energy $\Delta{}E$ between the nuclei and electrons.) The mechanism of the phenomenon is the inelastic scattering of the conduction electrons on the nucleus. Analogous process, i.e. the $^{229m}$Th$(e,e')^{229}$Th reaction, ensures fast decay of the $^{229m}$Th isomers inside a metal \cite{Tkalya-03}. Let $\Delta{}E = 10^{-6}$ eV. It corresponds to the photon wavelength $\lambda=2\pi/q\simeq0.5$ cm, where the transferred moment $q\simeq\Delta{}E\sqrt{m_e/2E_F}$ ($E_F=5.5$ eV is the Fermi energy, $m_e$ is the electron mass). At low temperature such photon is absorbed by a copper layer with depth $\delta<10^{-4}$ cm. The exchange by the virtual photon between the nuclei inside the crystal and the conduction electron in the metal cover take place because the effective ``mass'' of the photon $m_{\gamma}^*=\sqrt{{\bf{q}}^2-\Delta{}E^2}\simeq{}q$ is small. Such photon exists the time $\Delta{}t\simeq\hbar/m_{\gamma}^*$ and has the finite range $r_{\gamma}\simeq{}c\Delta{}t = 1/m_{\gamma}^*\simeq0.1$ cm. Since $r_{\gamma}\gg{}D + \delta$, the virtual photon emitted by the nucleus reaches the metal cover and interacts with the conduction electrons. This process can else be interpreted as internal conversion on the conduction electrons. It ensures the nuclear spin relaxation in $^{229}$Th:LiCaAlF$_6$.

The nuclear spin relaxation time is given by $1/T_1\simeq{}n_e(\Delta{}E/E_F)\sigma_e{}\xi_D{}v_F$,
where $n_e\simeq6\times10^{22}$ cm$^{-3}$ is the conduction electron density, $\sigma_e\simeq10^{-30}$ cm$^2$ is the $M1$ cross-section of the inelastic scattering of the Fermi energy electron on the $^{229}$Th ground state sublevels \cite{Alder-56}, $v_F=\sqrt{2E_F/m_e}$. The factor $\Delta{}E/E_F$ determines the number of electrons near the Fermi surface which can absorb or emit the quantum with the energy $\Delta{}E$ at low temperature. Nuclei are placed inside the dielectric sample with the diameter $D$. We take into account this fact by the factor $\xi_D=\sigma_e(D)/\sigma_e\simeq0.1$ ($\xi_D$ can be estimated through the radial integrals for the electron wave functions). Thus, the time required to achieve the Boltzman distribution population of the sublevels of the $^{229}$Th ground state is $\lesssim50$ d. It makes the third scheme of the $\gamma$-ray laser real. It is also worth noting that the metallic environment of the sample will not affect the 7.6 eV transition because $\lambda_{is}\ll{}D$.

And we close our consideration with a remark on the VUV lasers. The energy of the isomeric level is known roughly. Therefore we can not tell now, what VUV laser will be used for the pumping of the $^{229m}$Th isomers. It is possible that we can use one of the available lasers (see in \cite{Drake-06}). If not, it will be necessary to develop a special laser with the corresponding wavelength or use a free electron laser, which has a good tunability. For irradiation of the sample we need density of the photon flux $\varphi\simeq10^{20}$ cm$^{-2}$s$^{-1}$. Such $\varphi$ can be reached relatively easily by focusing of the radiation of middle power laser.

The author thanks Drs. A.~V.~Bibikov, I.~V.~Bodrenko, N.~N.~Delyagin, and A.~V.~Nikolaev for useful discussions.


\begin{thebibliography}{25}
\expandafter\ifx\csname natexlab\endcsname\relax\def\natexlab#1{#1}\fi
\expandafter\ifx\csname bibnamefont\endcsname\relax
  \def\bibnamefont#1{#1}\fi
\expandafter\ifx\csname bibfnamefont\endcsname\relax
  \def\bibfnamefont#1{#1}\fi
\expandafter\ifx\csname citenamefont\endcsname\relax
  \def\citenamefont#1{#1}\fi
\expandafter\ifx\csname url\endcsname\relax
  \def\url#1{\texttt{#1}}\fi
\expandafter\ifx\csname urlprefix\endcsname\relax\def\urlprefix{URL }\fi
\providecommand{\bibinfo}[2]{#2}
\providecommand{\eprint}[2][]{\url{#2}}

\bibitem[{\citenamefont{Baldwin and Solem}(1997)}]{Baldwin-97}
\bibinfo{author}{\bibfnamefont{G.~C.} \bibnamefont{Baldwin}} \bibnamefont{and}
  \bibinfo{author}{\bibfnamefont{J.~C.} \bibnamefont{Solem}},
  \bibinfo{journal}{Rev. Mod. Phys.} \textbf{\bibinfo{volume}{69}},
  \bibinfo{pages}{1085} (\bibinfo{year}{1997}).

\bibitem[{\citenamefont{Rivlin}(2007)}]{Rivlin-07}
\bibinfo{author}{\bibfnamefont{L.~A.} \bibnamefont{Rivlin}},
  \bibinfo{journal}{Quantum Electronics} \textbf{\bibinfo{volume}{37}},
  \bibinfo{pages}{723} (\bibinfo{year}{2007}).

\bibitem[{\citenamefont{Beck et~al.}(2007)}]{Beck-07}
\bibinfo{author}{\bibfnamefont{B.~R.} \bibnamefont{Beck}} \bibnamefont{et~al.},
  \bibinfo{journal}{Phys. Rev. Lett.} \textbf{\bibinfo{volume}{98}},
  \bibinfo{pages}{142501} (\bibinfo{year}{2007}).

\bibitem[{\citenamefont{Rellergert et~al.}(2010)}]{Rellergert-10}
\bibinfo{author}{\bibfnamefont{W.} \bibnamefont{Rellergert}}
  \bibnamefont{et~al.}, \bibinfo{journal}{Phys. Rev. Lett.}
  \textbf{\bibinfo{volume}{104}}, \bibinfo{pages}{200802}
  (\bibinfo{year}{2010}).

\bibitem[{\citenamefont{Tkalya}(2000)}]{Tkalya-00-JETPL}
\bibinfo{author}{\bibfnamefont{E.~V.} \bibnamefont{Tkalya}},
  \bibinfo{journal}{JETP Lett.} \textbf{\bibinfo{volume}{71}},
  \bibinfo{pages}{311} (\bibinfo{year}{2000}).

\bibitem[{\citenamefont{Tkalya et~al.}(2000)}]{Tkalya-00-PRC}
\bibinfo{author}{\bibfnamefont{E.~V.} \bibnamefont{Tkalya}}
  \bibnamefont{et~al.}, \bibinfo{journal}{Phys. Rev. C}
  \textbf{\bibinfo{volume}{61}}, \bibinfo{pages}{064308}
  (\bibinfo{year}{2000}).

\bibitem[{\citenamefont{Tkalya}(2003)}]{Tkalya-03}
\bibinfo{author}{\bibfnamefont{E.~V.} \bibnamefont{Tkalya}},
  \bibinfo{journal}{Physics-Uspekhi} \textbf{\bibinfo{volume}{46}},
  \bibinfo{pages}{315} (\bibinfo{year}{2003}).

\bibitem[{\citenamefont{Tkalya et~al.}(1996)}]{Tkalya-96}
\bibinfo{author}{\bibfnamefont{E.~V.} \bibnamefont{Tkalya}}
  \bibnamefont{et~al.}, \bibinfo{journal}{Phys. Scr.}
  \textbf{\bibinfo{volume}{53}}, \bibinfo{pages}{296} (\bibinfo{year}{1996}).

\bibitem[{\citenamefont{Peik and Tamm}(2000)}]{Peik-03}
\bibinfo{author}{\bibfnamefont{E.}~\bibnamefont{Peik}} \bibnamefont{and}
  \bibinfo{author}{\bibfnamefont{C.}~\bibnamefont{Tamm}},
  \bibinfo{journal}{Europhys. Lett.} \textbf{\bibinfo{volume}{61}},
  \bibinfo{pages}{181} (\bibinfo{year}{2000}).

\bibitem[{\citenamefont{Flambaum}(2006)}]{Flambaum-06}
\bibinfo{author}{\bibfnamefont{V.~V.} \bibnamefont{Flambaum}},
  \bibinfo{journal}{Phys. Rev. Lett.} \textbf{\bibinfo{volume}{97}},
  \bibinfo{pages}{092502} (\bibinfo{year}{2006}).

\bibitem[{\citenamefont{Dykhne and Tkalya}(1998{\natexlab{a}})}]{Dykhne-98}
\bibinfo{author}{\bibfnamefont{A.} \bibnamefont{Dykhne}} \bibnamefont{and}
  \bibinfo{author}{\bibfnamefont{E.} \bibnamefont{Tkalya}},
  \bibinfo{journal}{JETP Lett.} \textbf{\bibinfo{volume}{67}},
  \bibinfo{pages}{549} (\bibinfo{year}{1998}{\natexlab{a}}).

\bibitem[{\citenamefont{Dicke}(1954)}]{Dicke-54}
\bibinfo{author}{\bibfnamefont{R.~H.} \bibnamefont{Dicke}},
  \bibinfo{journal}{Phys. Rev.} \textbf{\bibinfo{volume}{93}},
  \bibinfo{pages}{99} (\bibinfo{year}{1954}).

\bibitem[{\citenamefont{Shiran et~al.}(2004)}]{Shiran-04}
\bibinfo{author}{\bibfnamefont{N.}~\bibnamefont{Shiran}} \bibnamefont{et~al.},
  \bibinfo{journal}{Radiation Measurement.} \textbf{\bibinfo{volume}{38}},
  \bibinfo{pages}{459} (\bibinfo{year}{2004}).

\bibitem[{\citenamefont{Bemis et~al.}(1988)}]{Bemis-88}
\bibinfo{author}{\bibfnamefont{C.~E.} \bibnamefont{Bemis}}
  \bibnamefont{et~al.}, \bibinfo{journal}{Phys. Scr.}
  \textbf{\bibinfo{volume}{38}}, \bibinfo{pages}{657} (\bibinfo{year}{1988}).

\bibitem[{\citenamefont{Barci et~al.}(2003)}]{Barci-03}
\bibinfo{author}{\bibfnamefont{V.}~\bibnamefont{Barci}} \bibnamefont{et~al.},
  \bibinfo{journal}{Phys. Rev. C} \textbf{\bibinfo{volume}{68}},
  \bibinfo{pages}{034329} (\bibinfo{year}{2003}).

\bibitem[{\citenamefont{Ruchowska et~al.}(2006)}]{Ruchowska-06}
\bibinfo{author}{\bibfnamefont{E.}~\bibnamefont{Ruchowska}}
  \bibnamefont{et~al.}, \bibinfo{journal}{Phys. Rev. C}
  \textbf{\bibinfo{volume}{73}}, \bibinfo{pages}{044326}
  (\bibinfo{year}{2006}).

\bibitem[{\citenamefont{Dykhne and Tkalya}(1998{\natexlab{b}})}]{Dykhne-98_ME}
\bibinfo{author}{\bibfnamefont{A.} \bibnamefont{Dykhne}} \bibnamefont{and}
  \bibinfo{author}{\bibfnamefont{E.} \bibnamefont{Tkalya}},
  \bibinfo{journal}{JETP Lett.} \textbf{\bibinfo{volume}{67}},
  \bibinfo{pages}{251} (\bibinfo{year}{1998}{\natexlab{b}}).

\bibitem[{\citenamefont{Strizhov and Tkalya}(1991)}]{Strizhov-91}
\bibinfo{author}{\bibfnamefont{V.~F.} \bibnamefont{Strizhov}} \bibnamefont{and}
  \bibinfo{author}{\bibfnamefont{E.~V.} \bibnamefont{Tkalya}},
  \bibinfo{journal}{Sov. Phys. JETP} \textbf{\bibinfo{volume}{72}},
  \bibinfo{pages}{387} (\bibinfo{year}{1991}).

\bibitem[{\citenamefont{Mossbauer}(1961)}]{Mossbauer-61}
\bibinfo{author}{\bibfnamefont{R.~L.} \bibnamefont{Mossbauer}},
  \bibinfo{journal}{Sov. Phys. Usp.} \textbf{\bibinfo{volume}{3}},
  \bibinfo{pages}{866} (\bibinfo{year}{1961}).

\bibitem[{\citenamefont{Abragam}(1961)}]{Abragam-61}
\bibinfo{author}{\bibfnamefont{A.}~\bibnamefont{Abragam}},
  \emph{\bibinfo{title}{The Principles of Nuclear Magnetism}}
  (\bibinfo{publisher}{Clarendon}, \bibinfo{address}{Oxford},
  \bibinfo{year}{1961}).

\bibitem[{\citenamefont{Amaral et~al.}(2003)}]{Amaral-03}
\bibinfo{author}{\bibfnamefont{J.} \bibnamefont{Amaral}}
  \bibnamefont{et~al.}, \bibinfo{journal}{J. Phys.: Cond. Matt.}
  \textbf{\bibinfo{volume}{15}}, \bibinfo{pages}{2523} (\bibinfo{year}{2003}).

\bibitem[{\citenamefont{Jackson et~al.}(2009)}]{Jackson-09}
\bibinfo{author}{\bibfnamefont{R.~A.} \bibnamefont{Jackson}}
  \bibnamefont{et~al.}, \bibinfo{journal}{J. Phys.: Cond. Matt.}
  \textbf{\bibinfo{volume}{21}}, \bibinfo{pages}{325403}
  (\bibinfo{year}{2009}).

\bibitem[{\citenamefont{Sen and Narasimhan}(1977)}]{Sen-77}
\bibinfo{author}{\bibfnamefont{K.} \bibnamefont{Sen}} \bibnamefont{and}
  \bibinfo{author}{\bibfnamefont{P.} \bibnamefont{Narasimhan}},
  \bibinfo{journal}{Phys. Rev. B} \textbf{\bibinfo{volume}{15}},
  \bibinfo{pages}{95} (\bibinfo{year}{1977}).

\bibitem[{\citenamefont{Alder et~al.}(1956)}]{Alder-56}
\bibinfo{author}{\bibfnamefont{K.} \bibnamefont{Alder}}
  \bibnamefont{et~al.}, \bibinfo{journal}{Rev. Mod. Phys.}
  \textbf{\bibinfo{volume}{28}}, \bibinfo{pages}{432}
  (\bibinfo{year}{1956}).

\bibitem[{Dra()}]{Drake-06}
\bibinfo{note}{{\it{Springer handbook of atomic, molecular, and optical
  physics}}. G.~W.~F.~Drake (Ed.), (Springer, 2nd ed., 2006).}

\end{thebibliography}
\end{document}